\documentclass[aps,twocolumn,showpacs]{revtex4}
\usepackage{graphicx}
\usepackage{epsfig}
\usepackage{epstopdf}
\usepackage{amsfonts}
\usepackage{amssymb}
\usepackage{amsbsy}
\usepackage{amsmath}
\usepackage{mathrsfs}
\usepackage{latexsym}
\usepackage{natbib}
\usepackage{bm}
\usepackage{ifpdf}

\usepackage{color}
\usepackage{braket}
\usepackage{slashed}
\usepackage{pgfplots}

\def\x{\mathbf{x}}

\def\k{\mathrm{k}}

\def\kfb{\bar{k}_F}

\begin{document}

\title{Reduced Chandrasekhar mass limit due to the fine-structure constant}

\author{Golam Mortuza Hossain}
\email{ghossain@iiserkol.ac.in}

\author{Susobhan Mandal}
\email{sm17rs045@iiserkol.ac.in}

\affiliation{ Department of Physical Sciences, 
Indian Institute of Science Education and Research Kolkata,
Mohanpur - 741 246, WB, India }
 

\date{\today}

\begin{abstract}

The electromagnetic interaction alters the Chandrasekhar mass limit by a factor 
which depends, as computed in the literature, on the atomic number of the 
positively charged nuclei present within the degenerate matter. Unfortunately, 
the methods employed for such computations break Lorentz invariance ab initio. 
By employing the methods of finite temperature relativistic quantum field 
theory, we show that in the leading order, the effect of electromagnetic 
interaction reduces the Chandrasekhar mass limit for non-general-relativistic, 
spherically symmetric white dwarfs by a universal factor of $(1-3\alpha/4\pi)$, 
$\alpha$ being the fine-structure constant.

\end{abstract}

\maketitle


\emph{Introduction.-- }
The first-ever detection of the gravitational waves \cite{Abbott:2016blz} has 
provided an unprecedented window to probe fundamental physics at a much deeper 
level. The recent observation of the gravitational waves from the merger of the 
binary neutron stars \cite{TheLIGOScientific:2017qsa} has also been accompanied 
by the electromagnetic observation of the same event. These combined 
observations, the so-called multi-messenger gravitational wave astronomy, has 
already began to put stringent constraint on the possible form of the equation 
state of the nuclear matter within the neutron stars 
\cite{Annala:2017llu,Abbott:2018exr, Nandi:2018ami}. The future detection of 
low-frequency gravitational waves \cite{lisa:amaro2012low} from the extreme 
mass-ratio merger of a black whole with a white dwarf could determine the 
equation of state of the degenerate matter within the white dwarf with an 
accuracy reaching up to $0.1\%$ \cite{Han:2017kre}. Such a high-precision 
measurement would imply a significant jump in accuracy in determining the 
equation of state of the white dwarfs over current astronomical measurements 
\cite{Holberg:2012pu,Tremblay:2017mnras2849,Magano:2017mqk} and would be able to 
test the expected corrections due to the electromagnetic interaction.

In the study of white dwarf physics, as pioneered by Chandrasekhar 
\cite{chandrasekhar1931maximum, chandrasekhar1935highly}, the effects of 
electromagnetic interaction \emph{i.e.} Coulomb effects on the 
equation of state were considered by Kothari \cite{kothari1938theory}, Auluck 
and Mathur \cite{auluck1959electrostatic} and later more accurately by Salpeter 
\cite{salpeter1961energy}. Usually these effects are considered by including the 
`classical' electrostatic energy of uniformly distributed degenerate electrons 
within Wigner-Seitz cells. Each of these primitive cells contains a positively 
charged nucleus at the center to make it overall charge neutral. Additionally, 
one considers the so-called Thomas-Fermi corrections which arise due to the 
radial variation of electron density within a Wigner-Seitz cell. Other 
corrections are obtained by considering the `exchange energy', the `correlation 
energy' of interacting electrons and relativistic corrections of Thomas-Fermi 
model \cite{rotondo:PhysRevD.84.084007}. These corrections modify the 
Chandrasekhar mass limit by a factor which depends on the atomic number of the 
positively charged nuclei \cite{hamada1961models,nauenberg1972analytic}.

On the other hand, the existence of the Chandrasekhar mass limit follows from 
the physics of special relativity. Therefore, the methods which rely on the 
electrostatic consideration to compute modifications to Chandrasekhar mass 
limit are not very reliable as they break Lorentz invariance \emph{ab initio}. A 
natural approach to compute the effects of electromagnetic interaction on the 
Chandrasekhar mass limit in a Lorentz invariant manner which also considers the 
fact that white dwarfs have finite temperature, would be to employ the methods 
of the finite temperature relativistic quantum field theory. Following the 
pioneering work of Matsubara \cite{matsubara1955new}, these techniques were used 
to compute the ground state energy of the relativistic electron gas including 
corrections due to the fine-structure constant in the context of \emph{quantum 
electrodynamics} (QED) by Akhiezer and Peletminskii \cite{akhiezer1960use}, and 
later by Freedman and McLerran \cite{PhysRevD.16.1147}. However, to describe the 
degenerate matter within white dwarfs, the action of QED alone is not 
sufficient as it does not describe the interaction between the degenerate 
electrons and the positively charged heavier nuclei which are usually bosonic 
degrees of freedom. We address this issue here by considering a Lorentz 
invariant interaction between the electrons and the positively charged nuclei.

In order to understand the scales of the system, let us consider a well known 
white dwarf Sirius B which has observed mass density $\rho \approx 2.8 \times 
10^6 ~gm/cc$ and the effective temperature $T \approx 25922~\mathrm{K}$ 
\cite{joyce:2018mnras481}. In \emph{natural units} (\emph{i.e.} Planck constant 
$\hbar$ and speed of light $c$ are set to unity), the corresponding temperature 
scale is $\beta^{-1} \equiv k_B T = 2.2~\mathrm{eV}$ whereas the associated 
Fermi momentum is $k_F = (3\pi^2 n_e)^{1/3} \approx 0.57 ~\mathrm{MeV}$ with 
$n_e$ being the number density of the degenerate electrons within the white 
dwarf. These two together then provide a key dimensionless parameter to 
characterize the white dwarf as
\begin{equation}\label{BetaKfRatio}
\beta k_F \approx 2.6 \times 10^5  ~.
\end{equation}
For different white dwarfs the parameter (\ref{BetaKfRatio}) varies between 
$10^4 - 10^7$. To describe the interior spacetime within white dwarfs 
we ignore the effects of general relativity and consider the spacetime to 
be described by the Minkowski metric $\eta_{\mu\nu}$. For spherically symmetric 
white dwarfs then it leads to the usual hydro-static equilibrium condition 
$dP/dr = - G M(r) \rho/r^2$ where $M(r)= \int_0^r 4\pi r'^2 \rho dr'$ denotes 
`the enclosed mass' within a radial distance $r$. $P$ and $\rho$ are pressure 
and mass density respectively.

\begin{figure}
\includegraphics[width=9cm]{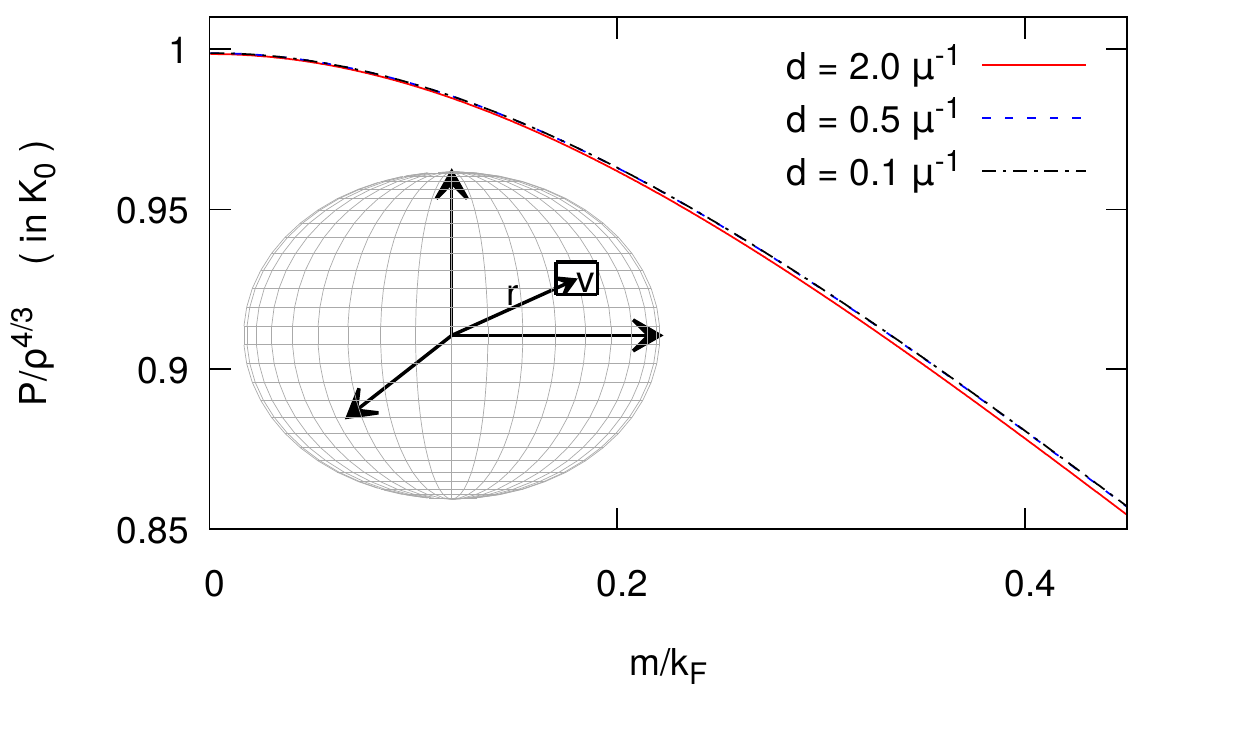}
\caption{(i) A finite box of volume $V$ at the given coordinate $r$ 
within a spherical star (inset figure). (ii) Different values of $d$ lead to 
same polytropic constant in the ultra-relativistic limit of the general 
equation of state.}
\label{fig:box-in-sphere}
\end{figure}

Within a spherically symmetric white dwarf, the pressure and mass density vary 
radially. However, to employ the techniques of finite temperature quantum field 
theory we have to consider a spatial region which is in thermal equilibrium at a 
given temperature $T$ along with uniform pressure and mass density. Therefore, 
around a given radial coordinate, we consider a finite spatial box which is 
sufficiently small so that the pressure and energy density can be treated to be 
uniform and yet sufficiently large to contain enough degrees of freedom to 
achieve required thermodynamical equilibrium (see FIG.\ref{fig:box-in-sphere}). 
The partition function that describes the degrees of freedom within the box, can 
be expressed as $\mathcal{Z} = Tr [e^{ -\beta (\hat{H} - \mu \hat{Q})}]$, 
where $\beta = 1/k_B T$ with $k_B$ being the Boltzmann constant, $\mu$ is the 
chemical potential and $Q$ refers to the conserved charge of the system. The 
Hamiltonian operator $\hat{H}$ represents the matter fields.


\emph{Matter fields.--}
The degenerate electrons within the box along with the spacetime metric 
$\eta_{\mu\nu} = diag(-1,1,1,1)$, are represented by the Dirac spinor field 
$\psi$ with the free-field action
\begin{equation}\label{FermionAction}
S_{\psi} = \int d^{4}x \mathcal{L}_{\psi}  = -\int d^{4}x  
~ \bar{\psi}[i \gamma^{\mu} \partial_{\mu} + m]\psi  ~,
\end{equation}
where the Dirac matrices $\gamma^{\mu}$ satisfies the anti-commutation relation
$\{\gamma^{\mu},\gamma^{\nu}\} = - 2 \eta^{\mu\nu} \mathbb{I}$. The minus 
sign in front of $\eta^{\mu\nu}$ is chosen such that the Dirac matrices 
satisfies the usual relations $(\gamma^0)^2 = \mathbb{I}$ and $(\gamma^k)^2 = 
-\mathbb{I}$ where $k=1,2,3$. %
The electromagnetic interaction between the electrons are mediated by the 
gauge fields $A_{\mu}$ and described by the action
\begin{equation}\label{InterctionAction}
S_{I}^{-} = \int d^4x \mathcal{L}_{I}^{-} = \int d^{4}x ~ 
\bar{\psi}[ e ~ \gamma^{\mu} A_{\mu} ]\psi ~,
\end{equation}
where $e$ is the electromagnetic coupling constant. On the other hand the 
free-field dynamics of $A_{\mu}$ is governed by the Maxwell action
\begin{equation}\label{MaxwellAction}
S_{A} = \int d^4x \mathcal{L}_{A} = \int d^{4}x ~ \left[ 
-\frac{1}{4}F_{\mu\nu } F^{\mu\nu} \right]  ~,
\end{equation}
where the field strength $F_{\mu\nu} = \partial_{\mu}A_{\nu} - 
\partial_{\nu}A_{\mu}$. The actions 
(\ref{FermionAction},\ref{InterctionAction},\ref{MaxwellAction}) together form 
the total action, say $S_{QED}$, used in the quantum electrodynamics.

The conserved 4-current corresponding to the action (\ref{FermionAction}) is 
given by $j^{\mu} = \bar{\psi} \gamma^{\mu} \psi$ which represents the 
contribution from the electrons. Similarly, we may consider a background 
4-current, say $J^{\mu}$, to represent the contribution from the positively 
charged nuclei which are usually bosonic degrees of freedom. Therefore, to 
describe the interaction between the electrons and the positively charged 
nuclei, here we consider a Lorentz invariant \emph{current-current} interaction 
term as follows
\begin{equation}\label{InteractionActionNucleon}
S_{I}^{+} = \int d^4x \mathcal{L}_{I}^{+} = \int d^{4}x  ~ 
[ - Z e^2 d^{2} ~ J_{\mu} \bar{\psi}\gamma^{\mu}\psi] ~.
\end{equation}
In Eq. (\ref{InteractionActionNucleon}), the coupling constant contains the 
term $- Z e^2 $ which signifies the strength of the attractive interaction 
between an electron and a positively charged nucleus with atomic number $Z$. The 
parameter $d$ which has the dimension of length, is introduced to make the 
action (\ref{InteractionActionNucleon}) dimensionless and it represents the 
interaction scale associated with the current-current interaction between the 
electrons and the nuclei. Therefore, the total action that describes the 
dynamics of the degenerate electrons within a white dwarf is given by
\begin{equation}\label{TotalMatterAction}
S = S_{QED} + S_{I}^{+} =  S_{\psi} + S_A + S_{I} =: \int d^4x 
\mathcal{L} ~,
\end{equation}
where  $S_{I} = S_{I}^{-} + S_{I}^{+}$. Inclusion of the additional interaction 
term (\ref{InteractionActionNucleon}) preserves the symmetry of the action 
$S_{QED}$. In other words, apart from being Lorentz invariant, the total action 
(\ref{TotalMatterAction}) is also invariant under local U(1) gauge 
transformations $\psi(x) \rightarrow e^{i\alpha(x)}\psi(x)$ and  
$A_{\mu}\rightarrow A_{\mu}-\frac{1}{e}\partial_{\mu}\alpha(x)$ with $\alpha(x)$ 
being an arbitrary function. Given the coupling constant $e$ is small, we can 
study the interacting theory by perturbative techniques of finite temperature 
quantum field theory.


\emph{Partition function.--}
To evaluate the partition function here we follow the path integral approach.  
In order to avoid over-counting of gauge degrees of freedom of $A_{\mu}$, 
it is convenient to introduce the Faddeev-Popov  ghost fields $C$ and $\bar{C}$ 
along with its action $S_C = \int d^4x \mathcal{L}_C = \int d^4x 
\partial^{\mu}\bar{C} \partial_{\mu}C$ \cite{book:17045,book:16435}. These 
Grassmann-valued fields effectively cancel the contributions from two gauge 
degrees of freedom. Therefore, the thermal partition function containing 
contributions from all the physical fields can be written as
\begin{equation}\label{TotalPartitionFunctionIntegral}
\mathcal{Z} = \int \mathcal{D}\bar{\psi} \mathcal{D}\psi
\mathcal{D}A_{\mu} \mathcal{D}\bar{C}~\mathcal{D}C ~ e^{-S^{\beta}} ~,
\end{equation}
where Euclidean action $S^{\beta} = \int_0^{\beta}d\tau \int d^3\x~ 
[\mathcal{L}+\mu\psi^{\dagger}\psi]_{|t=i\tau} =: S_{\psi}^{\beta} + 
(S_{A}^{\beta} + S_{C}^{\beta}) + S_{I}^{\beta}$ with $S_{\psi}^{\beta} = 
\int_0^{\beta}d\tau \int d^3\x [ \mathcal{L}_{\psi} + 
\mu\psi^{\dagger}\psi]_{|t=i\tau}$. We can express the total 
partition function using perturbative methods as $\ln\mathcal{Z} = 
\ln\mathcal{Z}_{\psi} + \ln\mathcal{Z}_A + \ln\mathcal{Z}_{I}$.

In the functional integral (\ref{TotalPartitionFunctionIntegral}) both fields
$A_{\mu}(x)$ and $C(x)$ are subject to the \emph{periodic} boundary conditions 
$A_{\mu}(\tau,\x) = A_{\mu}(\tau+\beta,\x)$ and $C(\tau,\x) = C(\tau+\beta,\x)$ 
whereas the spinor field is subject to the \emph{anti-periodic} boundary 
condition $\psi(\tau,\x) = -\psi(\tau+\beta,\x)$. The spinor field can be 
Fourier transformed as
\begin{equation}\label{FermionicFourier}
\psi(\tau,\x) = \frac{1}{\sqrt{V}} \sum_{n,\k} ~e^{i(\omega_n\tau + \k\cdot\x)} 
\tilde{\psi}(n,\k)  ~,
\end{equation}
where $V$ is the spatial volume of the box. The spinor field has mass dimension 
$3/2$ in natural units. So the Fourier modes $\tilde{\psi}(n,\k)$ are 
dimensionless. Further, the anti-periodic boundary condition implies that the 
Matsubara frequencies $\omega_n = (2n+1) \pi ~\beta^{-1}$ where $n$ is an 
integer. Using Eq. (\ref{FermionicFourier}), the Euclidean action for the spinor 
field can be expressed as 
\begin{equation}\label{PartitionFunctionSBEFourier}
S_{\psi}^{\beta} = \sum_{n,\k} ~\bar{\tilde{\psi}}~\beta
\left[ \slashed{p} - m \right] \tilde{\psi} ~,
\end{equation}
where $p_{\mu} = (p_0,\vec{p}) = (-i\omega_n+\mu,\k)$ and $\slashed{p} = 
\gamma^{\mu} p_{\mu}$. The Eq. (\ref{PartitionFunctionSBEFourier}) leads to 
momentum space thermal propagator for the free spinor field as 
$\mathcal{G}^{0}(\omega_n,\k) = 1/(\slashed{p} - m) = - (\slashed{p} + m)/(p^2 + 
m^2)$ where $p^2 = p^{\mu} p_{\mu}$. If one carries out the summation over $n$ 
by disregarding the formally divergent terms and the contribution from the 
anti-particles then fermionic part of the partition function becomes 
$\ln\mathcal{Z}_{\psi} = 2 \sum_{\k} \left[  \ln \left(1 + 
e^{-\beta(\omega-\mu)} \right) \right]$ where $\omega^2 = (\k^2 + m^2)$. The 
factor of 2 here denotes the spin-degeneracy of the electrons. To carry out the 
summation over $\k$, one may convert it to an integral as $\sum_{\k} \to V \int 
\frac{d^3\k}{(2\pi)^3}$. The Fermi momentum $k_F \equiv \sqrt{\mu^2-m^2}$ 
implies that for typical white dwarfs $\beta\mu \gg 1$. This strong inequality 
in turns allows the approximation $(e^{\beta(\omega - \mu)} + 1)^{-1} \simeq 
\Theta(\mu-\omega) - \mathrm{sgn}(\mu-\omega) e^{-\beta|\mu-\omega|}$ where  
$\Theta(\mu-\omega)$ is the Theta function and $\mathrm{sgn}(x)$ is the 
\emph{signum} function. The evaluation of the integral 
\cite{book:kapusta,Hossain:2019eos} then leads to
\begin{equation}\label{LogPartitionFunctionFermion1}
\ln\mathcal{Z}_{\psi} = \frac{\beta V}{24\pi^2} 
\left[2\mu k_F^3 - 3m^2 \kfb^2 + \frac{48 \mu k_F}{\beta^{2}} \right]  ~,
\end{equation}
where $\kfb^2 \equiv \mu k_F  - m^2 \ln \left((\mu + k_F )/m\right)$. The 
physical contribution from the gauge fields can be written as 
$\ln\mathcal{Z}_{A} = \ln(\int \mathcal{D}A_{\mu}\mathcal{D}\bar{C} 
~\mathcal{D}C ~e^{-(S_{A}^{\beta}+S_{C}^{\beta})}) = \tfrac{1}{45} V \pi^2 
\beta^{-3}$ which makes negligible contribution to the white dwarf equation of 
state and henceforth neglected.

The leading order contribution from the interaction terms can be expressed as 
$\ln\mathcal{Z}_{I} = \tfrac{1}{2} \langle (S_{-}^{\beta})^2 \rangle - \langle 
{S_{+}^{\beta}} \rangle$ where $\langle.\rangle$ denotes ensemble average.
The contribution due to the self-interaction of the electrons is 
\cite{book:kapusta,Hossain:2019eos}
\begin{equation}\label{SMinusSquaredAverageFourierSpaceFinal}
\langle (S_{-}^{\beta})^2 \rangle = \frac{\beta V e^2}{4\pi^2}
\left(\frac{\kfb^4}{4\pi^2} +  \frac{\kfb^2}{3\beta^2} \right) ~.
\end{equation}
Using the Eq. (\ref{InteractionActionNucleon}), we can express the contribution 
due to the interaction between the electrons and positively charged nuclei as
\begin{equation}\label{SPlusAverageRealSpace}
\langle S_{+}^{\beta} \rangle = -Z e^2 d^2 
\int_0^{\beta}d\tau \int d^3\x  J_{\mu}(\tau,\x) \langle
\overline{\psi}(\tau,\x)\gamma^{\mu}\psi(\tau,\x)\rangle ~.
\end{equation}
The Fourier space thermal propagator 
$\mathcal{G}(\omega_n,\k) = \int_0^{\beta}d\tau \int d^3\x 
~e^{-i(\omega_n\tau + \k\cdot\x)} \langle \psi(\tau_1,\x_1)  
\overline{\psi}(\tau_2,\x_2)\rangle$ along with  $\tau = \tau_1-\tau_2$, $\x = 
\x_1-\x_2$, leads the Eq. (\ref{SPlusAverageRealSpace}) to become
\begin{equation}\label{SPlusAverageFourierSpace}
\langle S_{+}^{\beta} \rangle = -Z e^2 d^2 \tilde{J}_{\mu}(\beta) \sum_{n,\k} ~
\mathrm{Tr} \left[\gamma^{\mu}\mathcal{G}(\omega_n,\k) \right] ~,
\end{equation}
where the trace is over the Dirac indices and the average background 
4-current density is $\tilde{J}_{\mu}(\beta) = 
(\beta V)^{-1} \int_0^{\beta}d\tau \int d^3\x ~ J_{\mu}(\tau,\x)$. We assume 
background 3-current density $\tilde{J}^k$ of the heavier nuclei is vanishing 
and identify corresponding charge density as $n_{+} \equiv \tilde{J}^0 = 
-\tilde{J}_0$. The Eq. (\ref{SPlusAverageFourierSpace}) then simplifies 
to
\begin{equation}\label{SPlusAverageFourierSpaceEvaluated}
\langle S_{+}^{\beta} \rangle = \frac{\beta V Z e^2 d^2 k_F^3 n_{+}}{3\pi^2} ~.
\end{equation}
Overall the system is electrically neutral. So the number density of positively 
charged nuclei must satisfy $Z n_+  = n_e$ where $n_e$ is the number density of 
the electrons. So the contribution to the partition function from the combined 
interaction becomes
\begin{equation}\label{LogPartitionFunctionInteractionTotal}
\ln\mathcal{Z}_{I} = \frac{\beta V e^2 }{96\pi^4}  
\left(3\kfb^4 - 32\pi^2 d n_e k_F^3 \right) ~,
\end{equation}
where we have ignored finite temperature corrections inside the parenthesis as 
the $(\beta k_F)^{-1}$ and coupling constant $e$ both are small.


\emph{Equation of state.--}
In order to understand the Chandrasekhar mass limit it is sufficient to evaluate 
the equation of state in its ultra-relativistic limit \emph{i.e.} when $k_F \gg 
m$, $\kfb \simeq k_F$ and $\mu \simeq k_F$ (see \cite{Hossain:2019eos} for 
general equation of state). As $(\beta k_F)^{-2} \sim 10^{-9}$ for typical white 
dwarfs then the Eqs. (\ref{LogPartitionFunctionFermion1}, 
\ref{LogPartitionFunctionInteractionTotal}) imply that the finite temperature 
corrections are much smaller compared to the corrections arising due to the 
fine-structure constant $\alpha \equiv e^2/4\pi \simeq 1/137$. Therefore, in the 
ultra-relativistic limit, the total partition function including the leading 
order $\alpha$ corrections but ignoring the finite temperature corrections, can 
be expressed as
\begin{equation}\label{LogPartitionFunctionTotalUR}
\ln\mathcal{Z} = \frac{\beta V}{12\pi^2} \left[k_F^4 + \frac{\alpha}{2\pi} 
\left(3 k_F^4 - 32 \pi^2 d^2 n_e k_F^3 \right) \right]  ~.
\end{equation}
The number density of the degenerate electrons can be computed as $n_e \equiv 
\langle N\rangle/V = (\beta V)^{-1} (\partial \ln \mathcal{Z}/\partial \mu)$.  
Given total partition function (\ref{LogPartitionFunctionInteractionTotal}) 
itself depends on the electron number density, it leads to an algebraic equation 
for $n_e$ as given below
\begin{equation}\label{ElectronNumberDensityFull0}
 n_e = \frac{k_F^3}{3\pi^2} \left[ 1 + \frac{3\alpha}{2\pi} 
 \left(1 - 8 \pi^2 d^2 n_e k_F^{-1}\right) \right] ~.
\end{equation}
The Eq. (\ref{ElectronNumberDensityFull0}) can be solved to express the 
corresponding mass density $\rho \equiv \mu_e m_u n_e$ as
\begin{equation}\label{MassDensityUR}
\rho = \frac{\mu_e m_u k_F^3}{3\pi^2} \left[ 1 + \frac{\alpha}{2\pi}
\left(3 - 2 d_{+}^2 \right) \right] ~,
\end{equation}
where $d_{+} \equiv 2d\mu$ and $m_u$ is the atomic mass unit. The chemical 
potential $\mu$ in the partition function provides a natural scale to construct 
the dimensionless parameter $d_{+}$ which characterizes the electron-nuclei 
interaction. The parameter $\mu_e \equiv (A/Z)$ with $A$ being the atomic 
mass number, is defined so that $\mu_e m_u$ specifies `the average mass per 
electron'.

In a grand canonical ensemble, we may read off the degeneracy pressure of 
the electrons as $P = (\beta V)^{-1}\ln \mathcal{Z}$ which leads to
\begin{equation}\label{PressureUR}
P = \frac{k_F^4}{12\pi^2} \left[ 1 + \frac{\alpha}{6\pi} 
\left(9  - 8 d_{+}^2\right) \right]  ~.
\end{equation}
By Combining the Eqs. (\ref{MassDensityUR}) and (\ref{PressureUR}), it is 
straightforward to write down a polytropic equation of state
$P = \mathrm{K} \rho^{4/3}$ with
\begin{equation}\label{EoSConstant}
\mathrm{K} = \frac{(3\pi^2)^{1/3} (1-\alpha/2\pi)}{4 (\mu_e m_u)^{4/3}}
= \mathrm{K}_0 (1-\alpha/2\pi) ~,
\end{equation}
where $K_0$ is the polytropic constant without $\alpha$ corrections. Clearly, 
the ratio $P/\rho^{4/3}$ for the degenerate matter becomes independent of the 
interaction between the electrons and the nuclei in the ultra-relativistic limit 
$k_F\gg m$ (see FIG.\ref{fig:box-in-sphere} for its dependence on $m/K_F$). 
However, in the non-relativistic limit, this interaction does contribute 
\cite{Hossain:2019eos}.


\emph{Chandrasekhar mass limit.--}
In order to find the Chandrasekhar mass limit, it is convenient to express the 
pressure as $P = \mathrm{K} \rho^{1+1/n}$ where $n=3$. Subsequently, one defines 
a dimensionless function $\theta(r)$ so that the mass density can be written as 
$\rho(r) = \rho_c \theta^n$. The identification of $\rho_c$ with central density 
implies $\theta(0) = 1$. The second boundary condition $\theta'(0) = 0$ follows 
from the condition $dP/dr = 0$ at $r=0$. Further, one defines a dimensionless 
variable $\xi = r/a$ where $a^2 = (\mathrm{K}/\pi G)\rho_c^{-2/3}$. The 
hydro-static equilibrium condition then leads to the Lane-Emden equation 
$\xi^{-2} \frac{d}{d\xi} \left(\xi^2 \frac{d\theta}{d\xi} \right) = - \theta^n$. 
 The Chandrasekhar mass limit is then defined as $M_{ch} = \int_0^R 4\pi r'^2 
\rho(r') dr'$ where $R$ is the radius of the white dwarf. As $n=3$ here, the 
Chandrasekhar mass limit can be explicitly expressed as $ M_{ch} = 
(4/\sqrt{\pi}) \left(\mathrm{K}/G\right)^{3/2} |\xi_0^2 ~\theta'(\xi_0)|$ where 
at the boundary $\xi_0 = R/a$ the mass density vanishes \emph{i.e.} 
$\theta(\xi_0) = 0$. The Lane-Emden equation can be solved numerically to find 
$|\xi_0^2 ~\theta'(\xi_0)| \simeq 2.02$. Therefore, including the \emph{leading} 
order effect of the fine structure constant, the Chandrasekhar mass limit 
becomes
\begin{equation}\label{ChandrasekharMassFSC}
M_{ch} = M_{ch}^0 \left(1 - \frac{3\alpha}{4\pi} \right)   ~,
\end{equation}
where $M_{ch}^0$ denotes the Chandrasekhar mass limit without $\alpha$ 
corrections. In other words, the effects of fine-structure constant reduces the 
Chandrasekhar mass limit by a universal factor which in the leading order does 
not depend on the atomic number $Z$ of the positively charge nuclei of the 
degenerate matter, unlike the results obtained in 
\cite{hamada1961models,nauenberg1972analytic}.

Using the value of the fine structure constant $\alpha \simeq 1/137$, we observe 
that the Chandrasekhar mass limit is reduced by $0.17 \%$ and similar order 
corrections are present in the corresponding equation of state. The future 
detection of low-frequency gravitational waves from the extreme mass-ratio 
merger of a black whole with a white dwarf could determine the equation of state 
of the degenerate matter within the white dwarf with an accuracy reaching up to 
$0.1\%$ \cite{Han:2017kre}. Therefore, the effects of fine-structure constant 
corrections as studied here would be within the detection threshold of such 
gravitational wave detectors in the future.

\emph{Acknowledgments.--} SM thanks CSIR, India for supporting this work 
through a doctoral fellowship.


\end{document}